# Quantum Fault Trees

Gabriel San Martín Silva[1, 2, a],
Tarannom Parhizkar[2, b] and Enrique López Droguett[1, 2, c]

[1] Department of Civil & Environmental Engineering, University of California, Los Angeles, USA.
[2] The B. John Garrick Institute for the Risk Sciences, University of California, Los Angeles, USA.
[a] gsanmartin@ucla.edu
[b] tparhizkar@ucla.edu
[c] eald@ucla.edu

**Abstract:** Fault tree analysis is a technique widely used in risk and reliability analysis of complex engineering systems given its deductive nature and relatively simple interpretation. In a fault tree, events are usually represented by a binary variable that indicates whether an event occurs or not, traditionally associated with the values 1 and 0, respectively. Different events are linked together using logical gates, modelling the dependencies that a subsystem or system may have over its basic components. In this study, quantum computing is leveraged to propose a novel approach to encode a traditional fault tree into a quantum algorithm. This quantum fault tree method uses quantum bits to represent basic events, effectively encoding the original fault tree into a quantum circuit. The execution of the resulting quantum circuit represents a full simulation of the fault tree, and multiple executions can be utilized to compute the failure probability of the whole system. The proposed approach is tested on a case study portraying a dynamic positioning system. Results verify that the quantum-based proposed approach is able to effectively obtain the dynamic positioning failure probability through simulation, opening promising opportunities for future investigations in the area.

## 1. INTRODUCTION

In recent years, early applications of quantum computing have been explored in different fields, such as finance ([1], [2]), medicine ([3], [4]) and material science [5], while promising significant speedups in the modeling and simulation processes. These promises are based on the leverage of quantum properties such as superposition, entanglement, and interference, which could be harnessed to perform certain tasks exponentially faster in terms of computational complexity [6], [7]. For the Reliability and Risk fields, early applications have been developed to tackle challenges such as condition-based monitoring using quantum neural networks [8] and structural health monitoring using Bayesian networks [9], [10]. In this study, a novel approach to embed a fault tree into a quantum circuit is introduced and discussed in detail.

Fault tree analysis (FTA) is a method for analyzing the failure probabilities of system states using Boolean logic. Fault trees are able to incorporate basic events interdependencies using logic block diagrams, including multiple events and gates such as AND/OR gates [11]. A detailed description of the concepts and techniques for FTs is presented in [12], [13]. In this paper, a case study of a fault tree for the control system of a dynamic positioning (DP) system is presented. The remaining of this paper is organized as follows. Section 2 presents the basic theory behind quantum computing, including the concepts of quantum bit and quantum gates. Section 3 introduces the proposed approach to encode a fault tree into a quantum algorithm, describing the translation of AND and OR gates into quantum circuits. Section 4 presents a case study in which a dynamic positioning system is analyzed using quantum fault trees and the results are compared to those obtained by a classical approach. Finally, Section 5 presents the concluding remarks of the paper.

## 2. THEORETICAL BACKGROUND

This section presents an overview of the main principles of quantum computing. In particular, the concepts of qubit and quantum gate are explained, drawing parallels to the traditional bit and logical gates when appropriate. Then, quantum circuits are presented to the reader as a framework to organize quantum operations and perform computation in a gate-based quantum computer or simulator. Finally, a brief exposition of the concept of measurement is introduced. These concepts form the required basis of knowledge to understand the proposed quantum fault tree approach, which is presented in section 3.

### 2.1. Qubits

A quantum bit, or qubit for short, is a generalization of the traditional concept of bit used thoroughly in modern computation. While both the bit and qubit represent two-states systems, the former is limited to only express one of two deterministic states at a time (either 0 or 1). The qubit, instead, can leverage the quantum property known as superposition to represent a linear combination of two mutually exclusive basis states, encoding a much more flexible representation. Mathematically, the qubit is an object residing in a two-dimensional Hilbert space, characterized by two complex parameters, as it is shown in equation (1):

$$|\psi\rangle = \alpha|0\rangle + \beta|1\rangle \qquad (1)$$

where the notation $|\psi\rangle$ is known as the ket notation and it is used in quantum mechanics to represent a vector [14]. In particular, $|0\rangle$ and $|1\rangle$ are the two basis vectors $[1 \ 0]^T$ and $[0 \ 1]^T$, respectively. Additionally, $\alpha$ and $\beta$ are complex numbers encoding probability amplitudes for each basis state, and therefore must fulfill the property shown in Equation (2):

$$|\alpha|^2 + |\beta|^2 = 1 \qquad (2)$$

It is often useful to represent a qubit in a graphical manner to visually interpret the operations that are performed to it. For this, a naïve approach would be to initially consider a 4-dimensional space, given that a qubit is fully characterized by two complex coefficients, each with two degrees of freedom. Nevertheless, it is possible to show that only two real-valued parameters are required to fully represent a qubit [14]. To see this, let us first write the expression for a qubit in its polar form, as shown in Equation (3):

$$|\psi\rangle = r_0 e^{i\phi_0}|0\rangle + r_1 e^{i\phi_1}|1\rangle \qquad (3)$$

where the four required parameters are identified as $\{r_0, r_1, \phi_0, \phi_1\}$, consisting of two amplitudes and two phases. Nonetheless, using Equation (2) it is possible to replace, without loss of generality, the parameters $r_0$ and $r_1$ by a single parameter $\theta$ using $r_0 = \cos\theta/2$ and $r_1 = \sin\theta/2$ (the use of the factor 1/2 will become evident later). Additionally, it is known that a quantum bit is not physically altered if amplified by a complex number of arbitrary phase but unitary norm [14]. Therefore, applying the new representation for the amplitudes $r_0$ and $r_1$, and multiplying the expression by $e^{-i\phi_0}$, equation (3) can be reduced to the expression portrayed in Equation (4):

$$|\psi\rangle = \cos\frac{\theta}{2}|0\rangle + \sin\frac{\theta}{2}e^{i(\phi_1-\phi_0)}|1\rangle \qquad (4)$$

Finally, denoting $\phi = \phi_1 - \phi_0$, the aforementioned expression is further reduced to:

$$|\psi\rangle = \cos\frac{\theta}{2}|0\rangle + \sin\frac{\theta}{2}e^{i\phi}|1\rangle \qquad (5)$$

where the representation of a qubit has been modified to show that only two real parameters, $\phi$ and $\theta$, are necessary to fully characterize it. Since these two parameters are often interpreted as angles, the qubit can also be understood as a surface point in a unitary sphere, where $\theta$ and $\phi$ are the polar and azimuthal angles, respectively. This sphere is called the Bloch Sphere, and it is shown in Figure 1. it is evident from Figure 1 that the factor $1/2$ in the angle $\theta$ is needed to ensure that the state $|1\rangle$ is obtained when $\theta = \pi$.

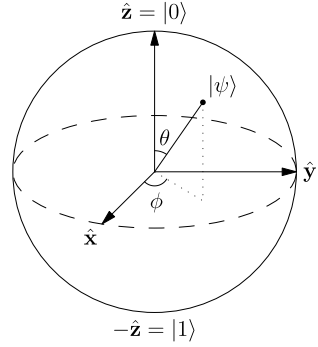

**Figure 1: Graphical representation of a qubit using the Bloch Sphere. Note how, for the cases where $\theta = 0$ and $\theta = \pi$, the qubit coincides with the basis states $|0\rangle$ and $|1\rangle$ respectively, as expected from Equation (5). Another special case is $\theta = \pi/2$, where the qubit occupies the equatorial zone of the Bloch sphere, representing a state of perfect superposition in which the probability of both basis states is the same and equal to $p_{|0\rangle} = p_{|1\rangle} = 0.5$.**

While more flexible than traditional bits, single qubits on their own cannot offer any significant computational advantage. It is in the interaction between different qubits where properties such as entanglement and interference arise and the advantages of quantum computing start to become relevant. $N$-qubit systems have the same representational power than $N$-bit systems: they can encode a total of $2^N$ possible states. The differentiating feature is that quantum systems composed of multiple qubits can encode, in the general case, all those states at the same time assigning different probability amplitudes to each of them. Mathematically, the state of a multiqubit system is computed from the individual qubit's states using the outer product, as shown in Equation (6):

$$|\Psi\rangle = |\psi_1\rangle \otimes |\psi_2\rangle \otimes \ldots \otimes |\psi_{N-1}\rangle \otimes |\psi_N\rangle \tag{6}$$

As an example, let us consider the case where $N = 2$, i.e., the multiqubit system is composed by two qubits: $|\psi_1\rangle = \alpha_1|0\rangle + \beta_1|1\rangle$ and $|\psi_2\rangle = \alpha_2|0\rangle + \beta_2|1\rangle$. The resulting state after performing the outer product operation is shown in Equation (7), where the definition for the ket vectors $|0\rangle$ and $|1\rangle$ has been used:

$$\begin{aligned}|\Psi\rangle &= |\psi_1\rangle \otimes |\psi_2\rangle \\ &= \left(\alpha_1\begin{bmatrix}1\\0\end{bmatrix} + \beta_1\begin{bmatrix}0\\1\end{bmatrix}\right) \otimes \left(\alpha_2\begin{bmatrix}1\\0\end{bmatrix} + \beta_2\begin{bmatrix}0\\1\end{bmatrix}\right) \\ &= \begin{bmatrix}\alpha_1\\\beta_1\end{bmatrix} \otimes \begin{bmatrix}\alpha_2\\\beta_2\end{bmatrix} \\ &= \begin{bmatrix}\alpha_1\begin{bmatrix}\alpha_2\\\beta_2\end{bmatrix}\\\beta_1\begin{bmatrix}\alpha_2\\\beta_2\end{bmatrix}\end{bmatrix} \\ &= \begin{bmatrix}\alpha_1\alpha_2\\\alpha_1\beta_2\\\beta_1\alpha_2\\\beta_1\beta_2\end{bmatrix}\end{aligned} \tag{7}$$

The resulting state can be further divided into a new set of basis vectors for a 4-dimensional Hilbert space, as depicted in Equation (8):

$$|\Psi\rangle = \begin{bmatrix} \alpha_1\alpha_2 \\ \alpha_1\beta_2 \\ \beta_1\alpha_2 \\ \beta_1\beta_2 \end{bmatrix}$$
$$= \alpha_1\alpha_2 \begin{bmatrix} 1 \\ 0 \\ 0 \\ 0 \end{bmatrix} + \alpha_1\beta_2 \begin{bmatrix} 0 \\ 1 \\ 0 \\ 0 \end{bmatrix} + \beta_1\alpha_2 \begin{bmatrix} 0 \\ 0 \\ 1 \\ 0 \end{bmatrix} + \beta_1\beta_2 \begin{bmatrix} 0 \\ 0 \\ 0 \\ 1 \end{bmatrix} \quad (8)$$
$$= \alpha_1\alpha_2|00\rangle + \alpha_1\beta_2|01\rangle + \beta_1\alpha_2|10\rangle + \beta_1\beta_2|11\rangle$$

where the basis vectors for the 4-dimensional Hilbert space have been also expressed in ket notation. As it can be seen, this multiqubit system is now able to encode four distinct probability amplitudes and, therefore, they must also fulfill the normalization condition described in Equation (2).

## 2.2. Quantum Gates

Quantum computation is based on the sequential modification of qubit states to perform a certain computational task. In gate-based computers, this modification is carried out by applying unitary matrices known as quantum gates to a single or multiple qubits. In this section, an overview of the quantum gates used for the proposed quantum fault trees is presented. For the sake of completeness, the Hadamard gate is included in the discussion even though it is not used in the case study presented in this work.

### 2.2.1 NOT Gate

The quantum NOT gate, also known as Pauli-X gate, is a single qubit gate that inverts the probability of both basis states. Mathematically, it is represented by the following matrix:

$$NOT = \begin{bmatrix} 0 & 1 \\ 1 & 0 \end{bmatrix} \quad (9)$$

### 2.2.2 Hadamard Gate

While the Hadamard gate is not used in the context of Quantum Fault Trees for the case study considered in this paper, it is presented here for the sake of completeness as it is one of the fundamental quantum computing gates. The Hadamard gate is also a single qubit gate whose main use is to induce a perfect superposition state in a qubit that was originally prepared to be in the $|0\rangle$ state. In matrix form, the Hadamard gate is represented as:

$$H = \frac{1}{\sqrt{2}} \begin{bmatrix} 1 & 1 \\ 1 & -1 \end{bmatrix} \quad (10)$$

### 2.2.3 Controlled-NOT (CX) Gate

While the Pauli-X and Hadamard gates are single qubit gates, the CX gate is a two-qubit gate, meaning that it operates over two qubits, generating quantum entanglement between them. The effect of the CX gate applied over a two-qubit system is the following: if the control qubit is in the $|1\rangle$ state, then a NOT gate is applied over the controlled qubit; otherwise, the gate has no effect. Mathematically, the CX gate is represented by the following matrix:

$$CX = \begin{bmatrix} 1 & 0 & 0 & 0 \\ 0 & 1 & 0 & 0 \\ 0 & 0 & 0 & 1 \\ 0 & 0 & 1 & 0 \end{bmatrix} \qquad (11)$$

While the traditional CX gate only has one control qubit, extensions to $N$ control qubits can be easily formulated. A special case is formed when $N = 2$, called the Toffoli Gate. As it will be shown in Section 3, the CX gate and its extensions gate are instrumental for generating the translation of the logical gates AND and OR into quantum circuits, allowing the encoding of fault trees into quantum computers.

**2.2.4 Y-Rotational Gate**

The aforementioned gates are non-parametric gates, meaning that they do not receive user-defined parameters to change the quantum state of the system. Rotational gates instead use external parameters to alter the probability amplitudes of the qubits within the system. The Y-Rotational gate, in particular, can be understood as executing a rotation with angle $\xi$ radians with respect to the Y-axis in the Bloch Sphere (Figure 1). This rotation effectively changes the probability amplitudes, augmenting the likelihood of measuring a $|0\rangle$ or $|1\rangle$ state, depending on the initial state and the angle $\xi$. Rotational gates represent a powerful way of interacting with quantum circuits, and they are the cornerstone of quantum machine learning methods such as Quantum Neural Networks [15]. As it will be seen in Section 3, the Y-rotational gate will be used to encode event-specific failure probabilities into the qubits of a system. The matrix representation of the Y-rotational gate is shown in Equation (12):

$$R_y(\xi) = \begin{bmatrix} \cos\frac{\xi}{2} & -\sin\frac{\xi}{2} \\ \sin\frac{\xi}{2} & \cos\frac{\xi}{2} \end{bmatrix}, \qquad (12)$$

**2.3. Quantum Circuits**

A quantum circuit can be defined as the ordered set of operations applied to a multi-qubit system. They represent the main approach to design algorithms in gate-based quantum computers. Every quantum circuit consists of at least three main steps. First, there is an initialization step, in which all the qubits are set into a specific state, usually the $|0\rangle$ state. Secondly, operations described as quantum gates are applied to the multi-qubit system in an orderly fashion, altering the probability amplitudes of the system, entangling different set of qubits. Finally, a measurement process is performed, collapsing the superposed quantum state of the system into one possible state according to the final probability amplitudes that were encoded into the circuit. This measurement process is a destructive operation in the sense that is irreversible: after a measurement is performed, the circuit ends as there is no way of reconfiguring the quantum states other than starting the circuit again. Due to the stochastic nature of the quantum system, each time the circuit is executed, the values recorded in the classical register will be different according to the underlying distribution specified by the probability amplitudes. Therefore, it is common practice to execute the circuit multiple times to obtain meaningful results

**3. PROPOSED APPROACH FOR QUANTUM FAULT TREES**

**3.1. General Overview of Fault Trees**

The quantum-based fault tree structure follows the same rules as a conventional fault tree, which are based on the application of logical gates to a set of basic events. After constructing a fault tree for a given system, a quantitative evaluation process is performed to compute the failure probability of the TOP event. Here, the equations for the conventional fault tree gates are presented, using as an example the simplified two event fault tree depicted in Figure 2.

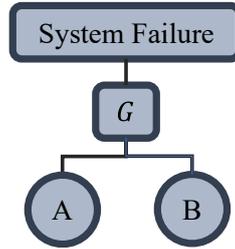

**Figure 2: A simplified fault tree with only two basic events, A and B. The final failure probability of the system will depend on the relationship between A and B, represented by the logical gate G.**

**AND gate:**
The system failure probability can be computed as:
$$P_f = P_A \cdot P_B \qquad (13)$$
where it is assumed that events A and B are independent.

**OR gate:**
For the OR gate, the system failure probability can be computed using De Morgan's theorem by inverting the gate, as shown in equation (14):
$$P_f = 1 - (1 - P_A) \cdot (1 - P_B) \qquad (14)$$

**NAND gate:**
Using the complement rule of probability, the system failure probability can be computed as:
$$P_f = 1 - P_A \cdot P_B \qquad (15)$$

**NOR gate:**
In a similar fashion as with the NAND gate, the system failure probability for the NOR gate is computed as:
$$P_f = (1 - P_A) \cdot (1 - P_B) \qquad (16)$$

For more complex fault trees, a similar process is performed in an iterative manner, starting from the bottom of the tree, and computing the failure probability of each subsystem. Then, the subsystem itself is considered as a basic event for higher level subsystems, until the TOP event is reached, and its probability is computed. A numerical example of this traditional approach is presented in Section 4 as the baseline to which que proposed Quantum Fault Tree will be compared against.

### 3.2. Quantum Logical Gates

The encoding of fault trees into a quantum circuit requires the translation of logical gates into their quantum equivalents. In this section, circuits corresponding to the application of an AND and OR logical gates over an arbitrary number of basic basics are presented and experimentally tested.

#### 3.2.1 Quantum AND Gate

The quantum AND gate for two inputs can be constructed using the Toffoli gate, as it is shown in Figure 3(a). Note how $q_a$ and $q_b$ are the control qubits and $q_c$ is the controlled qubit, which registers the result of the gate. Following the rules of the Toffoli gate, only when both $q_a$ and $q_b$ are in the $|1\rangle$ state, a NOT gate will be applied over $q_c$. If $q_c$ is initialized to be in the $|0\rangle$ state, then the Toffoli gate replicates exactly the truth table of a logical AND gate.

To extend the quantum AND gate for an arbitrary number of inputs, an N-Controlled NOT gate can be applied in a very similar fashion to the Toffoli gate: a NOT gate will be applied over $q_c$ only if $q_1$, $q_2$,…, $q_N$ are in the $|1\rangle$ state, changing $q_c$ from the original $|0\rangle$ state to the desired $|1\rangle$ state. A circuit implementing this behavior for three inputs is depicted in Figure 3(b).

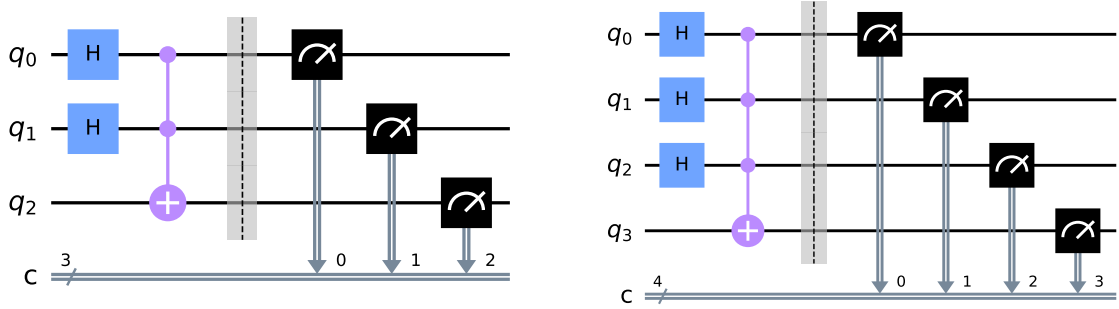

(a) Quantum AND gate with two inputs, $q_0$ and $q_1$, registering the output in $q_2$.

(b) Extension of the quantum AND gate for three inputs: $q_0$, $q_1$ and $q_2$.

**Figure 3:** Quantum AND gates with varying number of inputs. Note how in both (a) and (b) a Controlled-NOT sequence is used to determine the value of the output.

### 3.2.2 Quantum OR Gate

The quantum OR gate can be designed using De Morgan's theorem that indicates that a NAND gate is equivalent to an OR gate with inverted inputs. A NAND gate can be easily generated by inverting the output of the previously defined AND gate. The result is portrayed in Figure 4(a) as a circuit diagram for the case of two inputs. Note how NOT gates are applied two times to the inputs: once to invert them, before the application of the Toffoli gate, and once more after to return them to its original state. The output, instead, is only inverted once to transform the AND gate into a NAND gate. In a similar fashion as with the quantum AND gate, an extension of the OR gate is formulated by replacing the Toffoli gate by the more general N-controlled NOT gate, as depicted in Figure 4(b) for the case where three inputs are required.

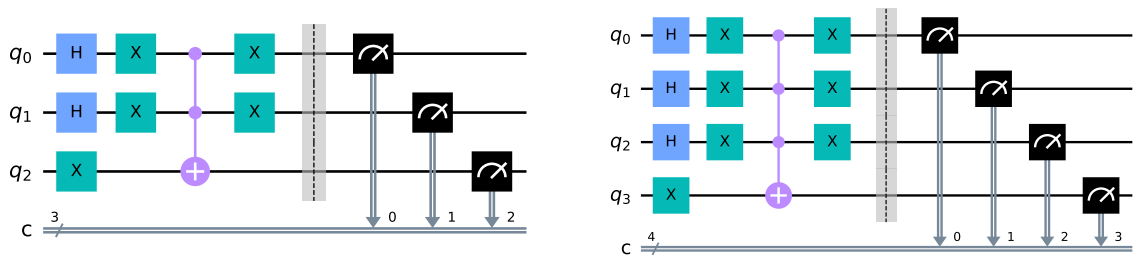

(a) Quantum OR gate with two inputs, $q_0$ and $q_1$, registering the output in $q_2$.

(b) Extension of the quantum OR gate for three inputs: $q_0$, $q_1$ and $q_2$.

**Figure 4:** Quantum OR gates with varying number of inputs. Note how in both (a) and (b) NOT gates are used before the Controlled-NOT sequence to invert the original AND gate, according to De Morgan's law. After the Controlled-NOT gate, another set of NOT gate is applied over the inputs to return them to their original state.

With quantum versions of the AND and OR gates designed, the encoding of a fault tree into a quantum circuit is explained in what follows.

### 3.3. Quantum Fault Trees

The proposed approach to encode a fault tree into a quantum circuit is comprised of the following set of steps:
1. Construct a quantum circuit with $N + M$ qubits, where $N$ is the number of basic events in the fault tree and $M$ is the number of quantum AND and OR gates that are required in the circuit.
2. Identify each basic event with a qubit. The basis states $|0\rangle$ and $|1\rangle$ will represent operational and failure states, respectively.

3. Apply a Y-Rotational gate to every basic event qubit to encode the failure probability of the event into the probability amplitudes of the qubit. Equation (17) [10] provides the angle required in the rotational gate to encode a failure probability $p_{|1\rangle}$:

$$\xi = 2 \cdot \tan^{-1}\left(\sqrt{\frac{p_{|1\rangle}}{1-p_{|1\rangle}}}\right) \qquad (17)$$

4. Starting from the bottom of the tree, encode each logical gate into the circuit using the quantum sub-circuits defined in Section 3.2. It is important to note that as the TOP event in the fault tree is reached, new gates will use the result from previous sub-systems as basic events.
5. Execute the circuit and record the measurement for every qubit into the classical register. This bitstring encodes a simulated scenario of the system, where each bit indicates whether that basic event, subsystem or system failed or not. The execution is repeated $K$ times to obtain multiple scenarios from which conclusions can be drawn.

Next, a case study depicting the fault tree of a dynamic positioning system and its translation into a quantum circuit is presented.

## 4. CASE STUDY: DYNAMIC POSITIONING SYSTEM

### 4.1. Description of the system

The case study considered in this paper presents the corresponding fault tree for the control system of a dynamic positioning (DP) system. As Figure 5 shows, the DP system controller can fail due to power blackout or a generalized failure in the computer system. These two subsystems are presented in the DP fault tree using an OR gate to indicate that both need to be in working condition for the DP system to operate correctly. The DP system has a redundant computer system with one unit in standby in case the main computer fails. That can be seen in the left side of the fault tree with two events named "Computer R1 fault" and "Computer R2 fault". When both computer units fail, then the main computer system fails; therefore, an AND gate is used to connect these events. Each computer unit can fail due to a hardware error, a software error, or a human error. Thus, the interconnection between these basic events for each computer is represented by an OR gate. The right side of the fault tree's diagram deals with the power system failure. For this particular DP system, a triple redundance was implemented, consisting of a main power generation unit and two uninterruptible power supplies (UPSs) as standby units. If all these three power sources fail, the DP system will fail in consequence. Therefore, in the fault tree, the power generation and UPSs failures are connected using an AND gate. Figure 5 presents a diagram of the fault tree for the dynamic positioning system. More detail about the control system failure model of DP system can be found in [16].

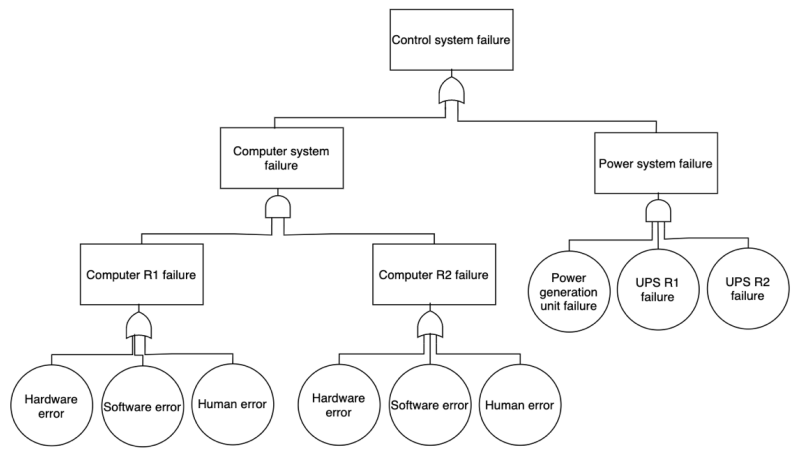

Figure 5: Fault tree for the case under study.

The failure probabilities for the basic events of the DP system are presented in Table 1 as specified in [16].

**Table 1: Annual failure probabilities for the basic events depicted in Figure 5**

| Basic Event | Annual Failure Probability |
|---|---|
| Hardware Error | $2.44 \times 10^{-5}$ |
| Software Error | $8.54 \times 10^{-5}$ |
| Human Error | $5.68 \times 10^{-1}$ |
| Power Generation Unit Failure | $3.41 \times 10^{-4}$ |
| UPS Failure | $3.66 \times 10^{-5}$ |

The failure probability of the TOP event (control system failure) is computed to generate a baseline to compare the results obtained by proposed quantum fault tree approach. For this, the process starts from the bottom of the fault tree by using the rules stablished in Section 3.1 to calculate the subsystem's failure probabilities. Given that the computer R1 and computer R2 are equivalent units, the annual failure probability is the same and is obtained as follows:

$$P_{Computer\ R_{1,2}\ failure} = 1 - (1 - 2.44 \times 10^{-5}) \cdot (1 - 8.54 \times 10^{-5}) \cdot (1 - 5.68 \times 10^{-1}) = 0.568 \quad (18)$$

Therefore, the computer system annual failure probability can be calculated as:

$$P_{Computer\ system\ failure} = 0.568 \cdot 0.568 = 0.323 \quad (19)$$

For the power generation branch, the annual failure probability for the power system is computed according to:

$$P_{Power\ system\ failure} = 3.41 \times 10^{-4} \cdot 3.66 \times 10^{-5} \cdot 3.66 \times 10^{-5} = 4.568 \times 10^{-13} \quad (20)$$

Finally, the control system's annual failure probability is:

$$P_{Power\ system\ failure} = 1 - (1 - 0.323) \cdot (1 - 4.568 \times 10^{-13}) = 0.323 \quad (21)$$

### 4.2. Implementation of the Quantum Fault Tree

Following the approach to construct a quantum fault tree described in Section 3.3, the DP system requires the implementation of a quantum circuit consisting of 14 qubits (i.e., 9 qubits to represent basic events and 5 qubits to implement the outputs of the AND and OR). Figure 6 depicts the quantum circuit that results from encoding the DP fault tree into a quantum circuit.

The circuit depicted in Figure 6 is executed 1,000,000 times in a desktop computer with an i5-7300HQ CPU and 16 GB of RAM, running Python 3.10.3 and the quantum simulator library Qiskit 0.19.2. The total execution time is 4.32 seconds. As previously explained, each execution samples from the underlying distribution encoded in the fault tree, effectively generating a simulation of the failure process. This simulation is extracted from the quantum circuit as a bitstring, in which each position indicates whether the correspondent component or subsystem failed or not, identified with the values 1 or 0.

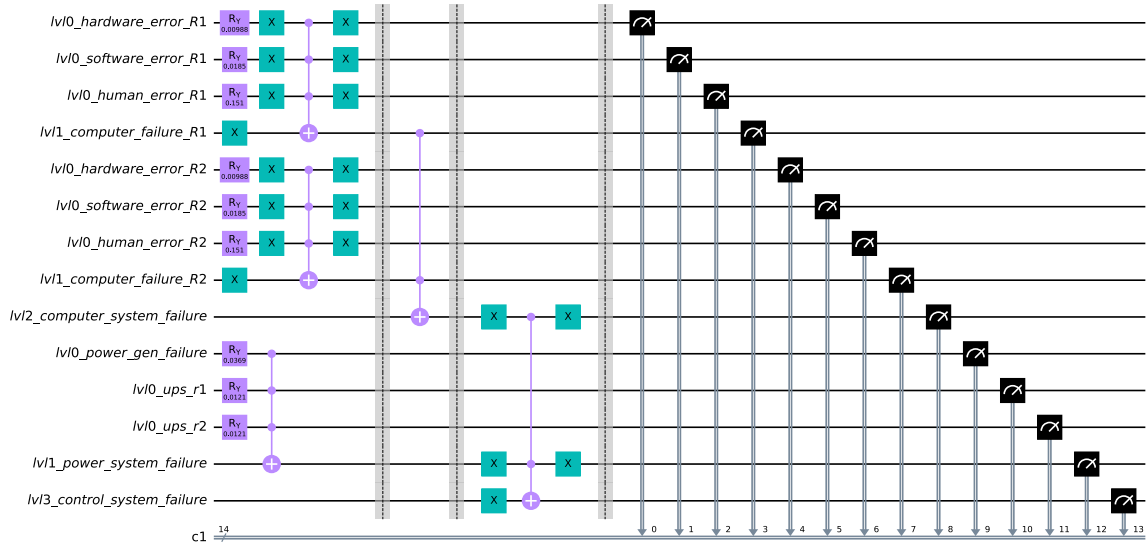

**Figure 6:** Dynamic positioning system encoded as a quantum circuit. Note how the probabilities are encoded into the qubits representing basic events using rotation-Y gates. The implementation of quantum AND and OR gates is executed starting from the bottom of the tree. Finally, the state of the system is identified as the result of measuring the last qubit, labelled as "lvl3_control_system_failure".

### 4.3. Results and Discussion

Figure 7 shows all the scenarios in which the DP control system failed, depicted as a histogram. Table 2 shows the equivalence between each bitstring and the final state of the system. As it can be seen, the case where the events "human error" occurs for both computers is by far the most common failure scenario. Other scenarios are significantly less frequent with differences of around three to four degrees of magnitude. This result can be explained by reviewing the probability failures shown in Table 1, where clearly human error is the most probable event to occur. As the failure of both computers automatically generates the failure of the whole system, human errors in both computers can be identified very quickly as a critical event using the proposed approach and, therefore, corrective action can be taken. While a similar conclusion can be reached using the traditional approach; for example, noticing from Equation (19) and Equation (20) that the failure probability of the computer system is much higher than for the power system, the proposed approach offers an alternative technique to identify weak points in the system. In this regard, the simulation process provided by the proposed quantum fault tree could be useful in quickly identifying hazardous events by analyzing the resulting bitstrings and their frequencies.

With respect to the control system failure probability, the total sum of scenarios in which the system fails in equal to 323,120. Given that the total number of scenarios simulated is 1,000,000, the failure probability of the control system can be numerically estimated to be 0.3231, in very close proximity to the one obtained by the traditional approach (in Equation (21)).

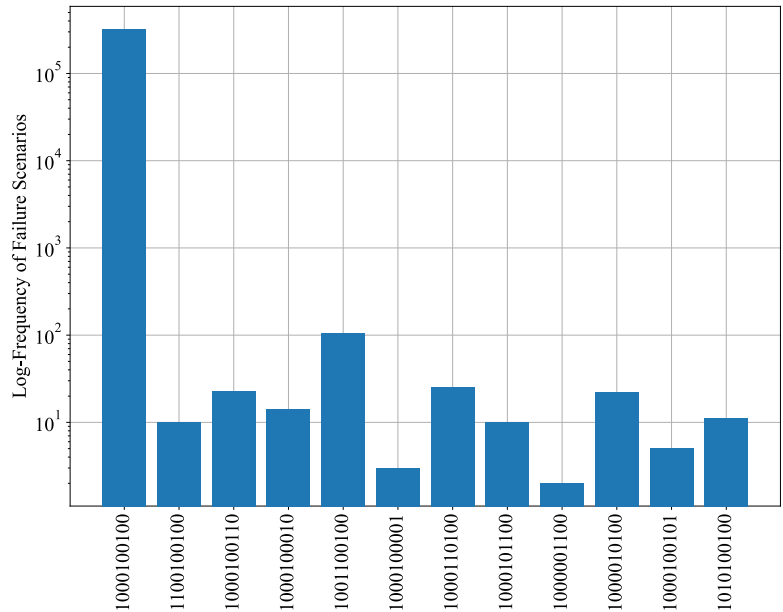

**Figure 7:** Frequency of simulated scenarios in logarithmic scale for every bitstring representing a failure in the control system. Note how every element contain a 1 in the first position, indicating that the state registered for the TOP event of the fault tree was "failed".

**Table 2:** Equivalence between bitstrings shown in Figure 7 and which components are in a failed state.

| Bitstring | Failed Components | Bitstring | Failed Components |
|---|---|---|---|
| 1000100100 | Human error in both computers (R1 and R2) | 1000110100 | Human error in both computers (R1 and R2) |
| 1100100100 | Human error in both computers (R1 and R2) and failure in UPS R1 | 1000101100 | Human error in both computers (R1 and R2) |
| 1000100110 | Human error in both computers (R1 and R2) and software error in computer R1 | 1000001100 | Human error in computer R1 and hardware error in computer R2 |
| 1000100100 | Human error in both computers (R1 and R2) | 1000010100 | Human error in computer R1 and software error in computer R2 |
| 1001100100 | Human error in both computers (R1 and R2) and failure in power generation unit. | 1000100101 | Human error in both computers (R1 and R2) and hardware error in computer R1 |
| 1000100001 | Human error in computer R2 and hardware error in computer R1 | 1010100100 | Failure in UPS R2, human error in both computers (R1 and R2) |

## 5. CONCLUDING REMARKS

This paper presents a novel approach to encode fault trees into a quantum computing algorithm to perform fault tree analysis and estimate the system's failure probability. For this, the AND and OR logical gates, commonly used in traditional fault tree analysis, are translated into equivalent quantum circuits. Additionally, rotational-Y quantum gates are used to embed failure probabilities into the probability amplitudes of quantum states, effectively achieving a one-to-one equivalence between qubits and basic events in the fault tree. The proposed approach is tested on a real case study, where the fault tree of a dynamic positioning system is presented.

Overall, it is found that the quantum fault tree approach is able to simulate different failure and non-failure scenarios through the execution of the underlying quantum circuit. From these simulations, summary statistics such as the system's failure probability can be estimated. Experimental results regarding the case study presented in this paper show that the results obtained using the proposed quantum Fault Tree approach match the results obtained through the analytical computation of the DP system's fault tree. Moreover, similar to traditional approaches, the collection of scenarios simulated using the quantum fault tree can also provide insights into the weak points of the system, allowing practitioners to take preventive action and allocate resources where they would be more efficient. As one of the main fields of study in the area is the sensitivity analysis of the results with respect to the number of circuit executions, scenario simulation based approaches using quantum computing could be a powerful alternative to efficiently estimate quantities of interests and understand how uncertainty is propagated in models.

Furthermore, it is the authors' opinion that this first approximation to a quantum-based fault tree model shows very promising results considering the relatively small number of qubits currently available in quantum simulation software. Nevertheless, the field of quantum computing is rapidly growing, with many companies developing both quantum hardware and software, and as such the early exploration of these techniques is relevant for the Probabilistic Risk Assessment community to maximize the benefits that might extracted when quantum hardware and algorithms provide real-world advantages over traditional approaches.